\documentstyle[11pt]{article}
\begin{document}
\vspace*{-2cm}
\noindent
\hspace*{11cm}
UG--FT--67/96 \\
\hspace*{11cm}
hep--ph/9607313 \\
\hspace*{11cm}
July 1996 \\
\begin{center}
\begin{large}
{\bf Numerical diagonalization of fermion mass matrices \\}
\end{large}
~\\
J. A. Aguilar--Saavedra \\
{\it Departamento de F\'{\i}sica Te\'{o}rica y del Cosmos \\
Universidad de Granada \\
18071 Granada, Spain}
\end{center}
\begin{abstract}
The diagonalization of general mass matrices is a more delicate problem when
eigenvalue degeneracies exist. In this case, often overlooked in the
literature, some difficulties arise related to the freedom in the choice of
basis in degenerate subspaces. Here two simple algorithms are developed to deal
with quark and neutrino mass matrices with arbitrary degeneracies.
\end{abstract}
\section{Introduction}
In a field theory with fermionic fields $\psi_i$, the mass sector of the
Lagrangian can be written as
\begin{equation}
-{\cal L}_{mass} = \bar \psi_i M_{ij} \psi_j ~+{\mathrm h.c.}
\label{ec:1}
\end{equation} 
The structure of the mass matrix $M$ is constrained by the symmetries of
the Lagrangian: charge conservation, baryon and lepton number conservation,
etc. We will consider the quark and neutrino sectors of the Standard Model
with $N$ generations of quarks; $n_L$ left-handed and $n_R$ 
right-handed
neutrinos. The mass terms in the quark sector are
\begin{equation}
-{\cal L}_{mass} = \bar u^0_{Li} M^u_{ij} u^0_{Rj} + \bar d^{\,0}_{Li} M^d_{ij} 
d^{\,0}_{Rj} + {\mathrm h.c.},
\label{ec:2}
\end{equation}
where $u^0_{L,R}$ ($d^{\,0}_{L,R}$) are the left and right-handed components of
the up
(down) quark fields, and $i,j=1,\dots,N$ label the different generations
of up (down) quarks. The
$N$-dimensional 
mass matrices $M^{u,d}$ are complex and arbitrary. In practice it is convenient
to work in a quark basis in which $M^{u,d}$ are diagonal. We  make an 
appropriate change of weak quark basis
\begin{equation}
u_{Li}=U^u_{Lij} u^0_{Lj}~,~~u_{Ri}=U^u_{Rij} u^0_{Rj}~,~~
d_{Li}=U^d_{Lij} d^{\,0}_{Lj}~,~~d_{Ri}=U^d_{Rij} d^{\,0}_{Rj}~,
\label{ec:3}
\end{equation}
such that the matrices $D^u \equiv U^u_L M^u U^{u\dagger}_R$, 
$D^d \equiv U^d_L M^d U^{d\dagger}_R$ are diagonal, and rewrite the mass term
as
\begin{equation}
-{\cal L}_{mass} = \bar u_{Li} D^u_{ij} u_{Rj} + \bar d_{Li} D^d_{ij} 
d_{Rj} + {\mathrm h.c.}
\label{ec:4}
\end{equation}
Of course this change of basis may affect other terms in the Lagrangian, but we
will not discuss this here. Thus what we denote here as `diagonalization' is 
really a biunitary transformation
\begin{equation}
U_1 M U^\dagger_2 = D
\label{ec:4b}
\end{equation}
such that $D$ is diagonal (we can also make the matrix elements of $D$ real and
positive). The matrices $U_{1,2}$ can be calculated \cite{papiro1} noting that
(\ref{ec:4b}) and $U_2 M^\dagger U^\dagger_1=D$ imply
\begin{equation}
U_1 M M^\dagger U^\dagger_1=D^2~,~~~U_2 M^\dagger M U^\dagger_2=D^2
\label{ec:5}
\end{equation}
However, this does not determine $U_{1,2}$ uniquely if $D$ has degenerate
diagonal elements, and this arbitrariness may cause that $D$ is no longer
diagonal: There are matrices $U_{1,2}$ satisfying (\ref{ec:5}) that do not
necessarily satisfy (\ref{ec:4b}). For instance, let us consider
\begin{equation}
M=\left(
\begin{array}{cc}
0 & 1 \\ 1 & 0
\end{array}
\right)
\label{ec:6}
\end{equation}
If we try to calculate $U_{1,2}$ by using Eqs. (\ref{ec:5}), we find $M
M^\dagger= M^\dagger M=I$, so $U_1=U_2=I$ and $U_1 M U^\dagger_2$ is not
diagonal. This simple example shows us that we need great care to deal with the
case of degenerate eigenvalues.

The mass terms in the neutrino sector are \cite{papiro3}
\begin{equation}
-{\mathcal L}_{mass}=\frac{1}{2} \bar \psi_L M_\nu \psi_R +
{\mathrm h.c.}
\label{ec:7}
\end{equation}
with
\begin{equation}
\psi_L=\left ( \begin{array}{c}
\nu_L \\ (\nu^c)_L
\end{array} \right )~,~~~
\psi_R=\left ( \begin{array}{c}
(\nu^c)_R \\ \nu_R
\end{array} \right )~,
\label{ec:8}
\end{equation}
and $M_\nu$ a complex $(n_L+n_R) \times (n_L+n_R)$ symmetric matrix. Under a
change of basis
\begin{equation}
N_L=U \psi_L~,~~N_R=U^* \psi_R
\label{ec:9}
\end{equation}
the mass term can be written as
\begin{equation}
-{\mathcal L}_{mass}=\frac{1}{2} \bar N_L D_\nu N_R +
{\mathrm h.c.}
\label{ec:10}
\end{equation}
with $D_\nu =U M_\nu U^T$ a diagonal matrix with real positive elements (it will
be shown below that any complex symmetric matrix $M_\nu$ can be `diagonalized'
with a congruent transformation of this type. A proof for the nondegenerate case
 can be found
in Ref. \cite{papiro4}). If $D$ is nondegenerate we can find the matrix $U$ by 
diagonalizing $M_\nu M_\nu^\dagger$, but if some degeneracies are present we can
get incorrect results.

In this paper we obtain two algorithms for the diagonalization of complex
arbitrary and complex symmetric matrices, respectively, and we write a package
{\tt Diagon} for their use with {\em Mathematica} 
\cite{papiro5}.\footnote{Although {\em Mathematica} has a built-in function 
{\tt SingularValues} that performs the diagonalization of an arbitrary matrix by
a biunitary transformation, it neglects all zero eigenvalues so it is useless 
for our purpose.} This package makes use of {\tt Eigensystem} to
calculate eigenvalues and eigenvectors of hermitian and unitary matrices.

\section{Algorithms}
We will first write an algorithm to diagonalize a complex arbitrary matrix by a
biunitary transformation and then use it to obtain a second one that performs 
the diagonalization of a complex symmetric matrix by a congruent transformation.
Note that we could use the first algorithm to diagonalize a complex symmetric 
matrix with two unitary matrices, but in general the transformation would not
be congruent.

Let $M$ be a complex arbitrary matrix. It is a well-known fact that we can find
two unitary matrices $U_1$, $U_2$ such that $U_1 M U_2^\dagger=D$ is a diagonal
matrix with real positive elements. We will assume for 
simplicity that $D$ has
diagonal elements $d_i$ with multiplicity $m_i$ and ordered from lower to
higher values. Let $V_1$, $V_2$ be
two matrices that diagonalize $M M^\dagger$ and $M^\dagger M$ respectively, 
$V_1 M M^\dagger V^\dagger_1=D^2$, $V_2 M^\dagger M V^\dagger_2=D^2$. It
can be proven that $V_i=K_i U_i$, where $K_i$ are block-diagonal unitary 
matrices that commute with $D$
\begin{equation}
D=\left(
\begin{array}{cccc}
d_0 I_{m_0} \\
& d_1 I_{m_1} \\
& & \ddots \\
& & & d_k I_{m_k}
\end{array}
\right)
~\Rightarrow~~
K_i=\left(
\begin{array}{cccc}
K_i^{(0)} \\
& K_i^{(1)} \\
& & \ddots \\
& & & K_i^{(k)} \\
\end{array}
\right)
\label{ec:11}
\end{equation}
(the dimension of the blocks $K_i^{(j)}$ is $m_j$). Then,
\begin{equation}
V_1 M V_2^\dagger=K_1 D K_2^\dagger \equiv D_B
\label{ec:12}
\end{equation}
The matrix in the right-hand side of Eq. (\ref{ec:12}) is a block-diagonal
matrix. Each block is an unitary matrix times the corresponding eigenvalue of
$D$: $D_B^{(j)}=d_j K_1^{(j)} K_2^{(j)\dagger}$. Furthermore, from $D_B
D_B^\dagger=D^2$ we can compute the eigenvalues $d_i$ and defining
$\lambda_i=1/d_i$ (if $d_0=0$, we take $\lambda_0=1$ and $D_B^{(0)}=I_{m_0}$),
\begin{equation}
\Lambda={\mathrm
diag}(\overbrace{\lambda_0,\dots,\lambda_0}^{m_0},\dots,
\overbrace{\lambda_k,\dots,\lambda_k}^{m_k})
\label{ec:13}
\end{equation}
we find that $U_1 \equiv V_1$, $U_2 \equiv \Lambda D_B V_2$ fulfill $U_1 M
U_2^\dagger=D$.

The diagonalization of a complex symmetric matrix $M_s$ by a congruent
transformation is somewhat more involved.
Let $U_1$, $U_2$ be two unitary matrices satisfying $U_1 M_s U_2^\dagger=D$, 
with
$D$ diagonal with real positive elements (we can calculate $U_{1,2}$ as it is
 shown
above). As $M_s$ is symmetric, $U_2^* M_s U_1^T=D$ and these equations imply
$U_2=K U_1^*$, with $K$ a block-diagonal unitary matrix that commutes with $D$
(analogous to $K_i$ in Eq. (\ref{ec:11})). Then,
\begin{equation}
U_1 M_s U_1^T=D K
\label{ec:14}
\end{equation}
We now want to decompose $K$ into two unitary symmetric matrices which commute
with $D$. The simplest way to show it is to see that $K$ is a symmetric matrix
(all the blocks $K^{(j)}$ are symmetric except possibly the block
corresponding to $d_0=0$, but in this case we can redefine $U_2$ without
changing $U_1$ and take $K^{(0)}=I$). A complex unitary symmetric matrix $K$ can 
always be
written as $K=e^{iS}$, with $S$ a real symmetric matrix \cite{papiro6}, and
$[K,D] =0$ implies $[ S,D] =0$, so we can rewrite Eq. (\ref{ec:14}) as
\begin{equation}
e^{-i \frac{S}{2}} U_1 M_s U_1^T e^{-i \frac{S}{2}}=D
\label{ec:15}
\end{equation}
However,  we
need a method to separate the matrix $K$ into two symmetric pieces.
First we diagonalize $K$ with a real orthogonal matrix $O$: $K=O^T K_d O$.
It can always be done\footnote{Note that $S$ is a real symmetric matrix} but we
must check that the subroutine we use to calculate
eigenvectors gives real results if it is possible (the built-in functions {\tt
Eigenvectors} and {\tt Eigensystem} do it).
We conventionally define the square root of $K$ as $K^{1/2}=O^T  K_d^{1/2} O$, 
with the
eigenvectors in the matrix $O$ conveniently ordered so that $K^{1/2}$ commutes
with $D$. Finally, $U=(K^{1/2})^\dagger U_1$ fulfills $U M_s U^T=D$.
\section{Examples}
Before we present two examples to illustrate the use of the package {\tt
Diagon}, it is important to notice that when the matrix {\tt m} has degenerate
eigenvalues, the set of eigenvectors calculated with {\tt u=Eigenvectors[m]} is
not necessarily an unitary matrix, and the product
{\tt u.Transpose[Conjugate[u]]} may have large off-diagonal elements 
$\epsilon \sim 10^{-1}$. This is not a roundoff error (typically of order
$O(10^{-16})$) as we show in the example below with an exact diagonalization.
We use {\tt DiagonalizeH}, a subroutine to diagonalize hermitian
matrices that uses {\tt Eigensystem} and the Gram-Schmidt method of
orthonormalization in each degenerate subspace to yield orthonormal
eigenvectors. The following example will illustrate this.
The numerical results have been obtained running {\em Mathematica} 
2.2.2 for Linux.

We write a previously calculated hermitian matrix {\tt m} with eigenvalues 
$(0,0,1,1)$ and diagonalize it. Note that {\tt DiagonalizeH} does not return the
eigenvectors but their complex conjugate.
\begin{verbatim}
In[1]:= <<Diagon.m
In[2]:= m=1/4 {{1,(1-I)/Sqrt[2],(-1+I)/Sqrt[2],1},{(1+I)/Sqrt[2],2,
        -1+I,(-1+I)/Sqrt[2]},{(-1-I)/Sqrt[2],-1-I,2,(-1+I)/Sqrt[2]},
        {1,(-1-I)/Sqrt[2],(-1-I)/Sqrt[2],3}};
In[3]:= u=Chop[Eigenvectors[m]];
In[4]:= u.Adj[u]
Out[4]= {{8, (-2 + 2 I) Sqrt[2], 0, 0}, {(-2 - 2 I) Sqrt[2], 4, 0, 0}, 
            8   2   2 I                    2   2 I           12
>    {0, 0, -, (- - ---) Sqrt[2]}, {0, 0, (- + ---) Sqrt[2], --}}
            5   5    5                     5    5            5
In[5]:= Simplify[Conjugate[u].m.Transpose[u]]
                                            8   2   2 I
Out[5]= {{0, 0, 0, 0}, {0, 0, 0, 0}, {0, 0, -, (- + ---) Sqrt[2]}, 
                                            5   5    5
             2   2 I           12
>    {0, 0, (- - ---) Sqrt[2], --}}
             5    5            5
In[6]:= u2=DiagonalizeH[m][[2]];
In[7]:= Chop[u2.Adj[u2]]
Out[7]= {{1., 0, 0, 0}, {0, 1., 0, 0}, {0, 0, 1., 0}, {0, 0, 0, 1.}}
In[8]:= Chop[u2.N[m].Adj[u2]]
Out[8]= {{0, 0, 0, 0}, {0, 0, 0, 0}, {0, 0, 1., 0}, {0, 0, 0, 1.}}
\end{verbatim}
The package {\tt Diagon} (see the {\em Appendix} for a complete 
description) contains the subroutines {\tt DiagonalizeH}, {\tt 
Diagonalize} and {\tt DiagonalizeS} . It has been tested more than 100 times on
{\em Mathematica}
2.2.2 for Linux and {\em Mathematica} 2.2 for Windows.
We have used the 
instructions listed below with different sets of
eigenvalues, obtaining always correct results. First we load the
package and define the matrices {\tt m} and {\tt ms} to test {\tt Diagonalize}
and {\tt DiagonalizeS} respectively. We use 16-digit precision in the input
matrices.
\begin{verbatim}
In[9]:= SeedRandom[]
In[10]:= m1=Table[Random[Complex,{-1-I,1+I}],{i,8},{j,8}];
In[11]:= m2=Table[Random[Complex,{-1-I,1+I}],{i,8},{j,8}];
In[12]:= ma1=m1+Adj[m1];
In[13]:= ma2=m2+Adj[m2];
In[14]:= mu1=Chop[Eigenvectors[ma1]];
In[15]:= mu2=Chop[Eigenvectors[ma2]];
In[16]:= m=mu1.DiagonalMatrix[{0,0,1,1,1,1,2,2}].mu2;
In[17]:= ms=mu1.DiagonalMatrix[{0,0,1,1,1,1,2,2}].Transpose[mu1];
In[18]:= sol=Diagonalize[m];
In[19]:= sol[[1]]
Out[19]= {0, 0, 1., 1., 1., 1., 2., 2.}
In[20]:= Chop[sol[[2]].m.Adj[sol[[3]]]]
Out[20]= {{0, 0, 0, 0, 0, 0, 0, 0}, {0, 0, 0, 0, 0, 0, 0, 0}, 
>    {0, 0, 1., 0, 0, 0, 0, 0}, {0, 0, 0, 1., 0, 0, 0, 0}, 
>    {0, 0, 0, 0, 1., 0, 0, 0}, {0, 0, 0, 0, 0, 1., 0, 0}, 
>    {0, 0, 0, 0, 0, 0, 2., 0}, {0, 0, 0, 0, 0, 0, 0, 2.}}
In[21]:= sol2=DiagonalizeS[ms];
In[22]:= sol2[[1]]
Out[22]= {0, 0, 1., 1., 1., 1., 2., 2.}
In[23]:= Chop[sol2[[2]].ms.Transpose[sol2[[2]]]]
Out[23]= {{0, 0, 0, 0, 0, 0, 0, 0}, {0, 0, 0, 0, 0, 0, 0, 0}, 
>    {0, 0, 1., 0, 0, 0, 0, 0}, {0, 0, 0, 1., 0, 0, 0, 0}, 
>    {0, 0, 0, 0, 1., 0, 0, 0}, {0, 0, 0, 0, 0, 1., 0, 0}, 
>    {0, 0, 0, 0, 0, 0, 2., 0}, {0, 0, 0, 0, 0, 0, 0, 2.}}
\end{verbatim}

\vspace{1cm}
\noindent
{\Large \bf Acknowledgements}

\vspace{0.4cm}
\noindent
I wish to thank F. del Aguila and M. Masip for revising the manuscript. I
also thank J. Jim\'{e}nez for useful comments. This work
was partially supported by CICYT under contract AEN94-0936, by the Junta de 
Andaluc\'{\i}a and by the European Union under contract CHRX-CT92-0004. 

\appendix
\newpage
\section{Appendix}
\begin{verbatim}
(* Title: Diagon *)

(* Author: J. A. Aguilar-Saavedra *)

(* Summary:
This package provides some algorithms for the diagonalization
of complex arbitrary and complex symmetric matrices
*)

(* Version: 1.1 *)

(* 
This package is available by anonymous ftp at deneb.ugr.es
in directory pub/packages/
*)

BeginPackage["Diagon`","LinearAlgebra`Orthogonalization`"]

Adj::usage =
 "Adj[m] gives Transpose[Conjugate[m]]"

Ortho::usage =
 "Ortho[{v,u}] takes a list of eigenvalues v and eigenvectors u and
 returns a list {v,u'} consisting of the eigenvalues v and 
 orthonormal eigenvectors u'"
 
DiagonalizeH::usage =
 "DiagonalizeH[m] gives a list {v,u}, with the eigenvalues and 
 eigenvectors of the square matrix m, such that 
 u.m.Transpose[Conjugate[u]]=DiagonalMatrix[v]. The list of 
 eigenvalues v is ordered by increasing value"

Diagonalize::usage =
 "Diagonalize[m] gives a list {v,u1,u2} where v is a row vector, and
  u1,u2 are matrices fulfilling and u1.m.Transpose[Conjugate[u2]]=
  DiagonalMatrix[v]"

DiagonalizeS::usage =
 "DiagonalizeS[m] gives a list {v,u} where v is a row vector and u is
  a matrix fulfilling u.m.Transpose[u]=DiagonalMatrix[v]"
 
Begin["`Private`"]

(* Definitions for internal use only *)

(* Increase the dimension of a matrix with the identity *)

IncDim1[m_]:=Transpose[Prepend[Transpose[Prepend[m,Table[0,
{Length[m]}]]],Join[{1},Table[0,{Length[m]}]]]]

IncDim[m_,nn_]:=Nest[IncDim1,m,nn]

(* Pick up a submatrix *)

DecDim[m_,nn_]:=Drop[Transpose[Drop[Transpose[m],nn]],nn]

(* Main definitions *)

Adj[m_]:=Transpose[Conjugate[m]];


(* Note that the scalar product that works is linear in the
   first variable and antilinear in the second:
   InnerProduct->(Dot[Conjugate[#2],#1]&)  *)

Ortho[sys_]:=Block[{i,j,allvalues,vectors,values,orthvectors},(
allvalues=N[Chop[sys[[1]]]];
vectors=N[Chop[sys[[2]]]];
For[i=1,i<=Length[allvalues],i++,
For[j=i+1,j<=Length[allvalues],j++,
If[Chop[allvalues[[i]]-allvalues[[j]]]==0,
allvalues[[j]]=allvalues[[i]] ] ] ];
values=Union[allvalues];
orthvectors=Flatten[Table[GramSchmidt[vectors[[Flatten[
Position[allvalues,values[[i]]]]]],
InnerProduct->(Dot[Conjugate[#2],#1]&)],{i,Length[values]}],1];
{Sort[allvalues],Chop[orthvectors]}
)];

DiagonalizeH[m_]:=Conjugate[
Ortho[Eigensystem[N[(m+Adj[m])/2] ] ] ]/;MatrixQ[m] &&
Length[m]==Length[Transpose[m]]

Diagonalize[m_]:=Block[{i,sol1,sol2,v,nz,mbox0,mbox,mod,mod2},(
sol1=DiagonalizeH[m.Adj[m]];
sol2=DiagonalizeH[Adj[m].m];
v=Chop[Sqrt[sol1[[1]]]];
nz=Count[v,0];
mbox0=Chop[sol1[[2]].m.Adj[sol2[[2]]]];
mbox=DecDim[mbox0,nz];
mod=Chop[mbox.Transpose[Conjugate[mbox]]];
mod2=DiagonalMatrix[Table[1/Abs[Sqrt[mod[[i,i]]]],{i,Length[mod]}]];
{v,Chop[sol1[[2]]],Chop[IncDim[mbox.mod2,nz].sol2[[2]]]}
)]/;MatrixQ[m] && Length[m]==Length[Transpose[m]] &&
Complement[Flatten[Chop[m]],{0}]!={}

Diagonalize[m_]:={Table[0,{Length[m]}],IdentityMatrix[Length[m]],
IdentityMatrix[Length[m]]}/;MatrixQ[m] &&
Length[m]==Length[Transpose[m]] && 
Complement[Flatten[Chop[m]],{0}]=={}

DiagonalizeS[m_]:=Block[{sol,v,u1,u2,nz,k0,k,sol2,rot,kd},(
sol=Diagonalize[m];
v=Chop[sol[[1]]];
u1=sol[[2]];
u2=sol[[3]];
nz=Count[v,0];
k0=Chop[Conjugate[u2].Adj[u1]];
k=DecDim[k0,nz];
sol2=Ortho[Eigensystem[k]];
rot=Sort[Conjugate[sol2[[2]]],OrderedQ[{Abs[#2],Abs[#1]}]& ];
kd=Chop[rot.k.Adj[rot]];
{v,Chop[IncDim[Adj[rot].Sqrt[kd].rot,nz].u1]}
)]/;MatrixQ[m] && Length[m]==Length[Transpose[m]] &&
Complement[Flatten[Chop[m]],{0}]!={}

DiagonalizeS[m_]:={Table[0,{Length[m]}],
IdentityMatrix[Length[m]]}/;MatrixQ[m] &&
Length[m]==Length[Transpose[m]] &&
Complement[Flatten[Chop[m]],{0}]=={}

End[]

EndPackage[]
\end{verbatim}

\end{document}